\documentclass[pdflatex,sn-basic,iicol]{sn-jnl}

\usepackage{amsmath}
\usepackage{amssymb}
\usepackage{bm}
\usepackage{mathrsfs}
\usepackage{subcaption}
\usepackage{booktabs, multirow} 

\jyear{2024}%

\theoremstyle{thmstyleone}%
\theoremstyle{thmstyletwo}%

\theoremstyle{thmstylethree}%

\begin{document}

\title[Shell topology optimization]{Shell topology optimization based on level set method}


\author*[1]{\fnm{Hiroki} \sur{Kobayashi}}\email{hiroki.kobayashi@mosk.tytlabs.co.jp}
\author[1]{\fnm{Katsuya} \sur{Nomura}}
\equalcont{Present address: \orgname{School of Engineering, Kwansei Gakuin University}, \orgaddress{\street{1 Gakuen Uegahara}, \city{Sanda}, \postcode{669-1330}, \state{Hyogo}, \country{Japan}}}
\author[2]{\fnm{Yuqing} \sur{Zhou}}
\author[2,1]{\fnm{Masato} \sur{Tanaka}}
\author[1]{\fnm{Atsushi} \sur{Kawamoto}}
\author[1]{\fnm{Tsuyoshi} \sur{Nomura}}

\affil[1]{
\orgname{Toyota Central R\&D Labs., Inc.}, \orgaddress{\street{1-4-14 Koraku}, \city{Bunkyo-ku}, \postcode{112-0004}, \state{Tokyo}, \country{Japan}}}

\affil[2]{
\orgname{Toyota Research Institute of North America, Toyota Motor North America}, \orgaddress{\street{1555 Woodridge Ave.}, \city{Ann Arbor}, \postcode{48105}, \state{MI}, \country{USA}}}



\abstract{
This paper proposes a level set-based method for optimizing shell structures with large design changes in shape and topology. Conventional shell optimization methods, whether parametric or nonparametric, often only allow limited design changes in shape. In the proposed method, the shell structure is defined as the isosurface of a level set function. The level set function is iteratively updated based on the shape sensitivity on the surface mesh. Therefore, the proposed method can represent an arbitrary manifold surface while dealing with topological changes, for example, from a spherical surface to a toroidal surface. We applied the proposed method to the mean compliance minimization problems of 3D shell structural designs for dome, bending plate and cantilever beam examples to demonstrate its efficacy of the proposed method. 
}




\maketitle

\section{Introduction}
Shell structures are used for a variety of industrial purposes including automobiles, aircraft, and buildings, because of their ease of economic mass production and lightweight. A thin curved shell, which is stiff against in-plane forces can be easily deformed against out-of-plane bending. Such thin shell structures must be appropriately designed to avoid large bending while effectively supporting the prescribed loads, which is difficult to achieve using only intuitive or empirical methods. For problems involving such challenges, structural optimization can be a powerful method for a wide variety of problems. 

In structural optimization methods, numerical analysis methods, such as finite element method (FEM), play an important role in accurately analyzing the performance of designs. In terms of mesh, analysis of the shells can be divided into two main types: volume mesh and surface mesh. 

The volume mesh-based methods, which discretize an analysis domain into finite volume elements, are widely used in structural optimization. 
Specifically, the topology optimization method has the advantage of optimizing structures with a high degree of freedom. 
After the pioneering work by \citet{bendsoe1988generating}, topology optimization has been applied to various applications in the structural mechanics field, for example, aeroplane wings~\citep{aage2017giga}, automotive components~\citep{kim2020topology}, and drone frames made of composites~\citep{zhou2022large}. 

The basic concept of topology optimization is to replace the original structural optimization problem by a material distribution problem in a fixed design domain such that arbitrary shapes can be represented. Usually, the topology optimization results in nonuniform thickness, but in the case of shell structures, they must have a certain plate thickness due to the raw material and manufacturing constraints. Therefore, topology optimization methods with geometric constraints have been studied extensively. 

\citet{guest2009imposing} introduced the maximum length scale control in topology optimization. \citet{allaire2016thickness} proposed a framework to constrain the local thickness via a level set-based method. \citet{zhou2022topology} proposed a method for optimizing 3D thin-walled structures through two PDE-based filtering operations and an aggregation constraint. Such methods can be used to optimize shell-like structures while maintaining a near-uniform thickness. A coating approach~\citep{clausen2015topology}, which represents thin structures in the transition region between void and solid (infill), has been used for shell structure optimization. A level set-based method has also been applied for coated structures \citep{wang2018level}. In their work, the coating material is represented by a zero-neighborhood of the level set function. The coating approach has mainly been applied to the shells with infills such as porous infill~\citep{wang2018design} or graded lattice~\citep{wu2017minimum}. \citet{clausen2017topology} discussed assigning void to the interior, representing pure shell structures. 

However, the volume mesh-based methods are computationally challenging. To represent a shell structure in volume mesh, its analysis domain must be discretized sufficiently in the thickness direction. Therefore, the resolution of the mesh is determined by the thickness of the shell. For example, if a shell with the thickness of 1/100 in a $1 \times 1 \times 1$ design domain is discretized to five elements for the thickness, the total number of elements reaches to $500^3 = 125$M in uniform cubic elements. Adaptive meshing approaches, such as \citet{li2021full} and \citet{zhou2022topology}, can reduce the number of elements. Despite these techniques, it is still computationally expensive because the mesh size in the shell region has to be very small to maintain certain elements in thickness with the proper aspect ratio. 

On the other hand, the surface mesh-based methods are computationally efficient because it requires less elements. The surface mesh discretize an analysis domain into 2D elements, i.e., elements without thickness. The thickness of the shell is considered in a mathematical model~\citep{chapelle2010finite}. Since the surface mesh does not need to be discretized in the thickness direction, even very thin structures can be analyzed with a few elements. 

For the structural optimization using the surface mesh, shape optimization methods have been well studied. Shape optimization methods can be classified into parametric and non-parametric methods. The parametric methods represent the shapes by predetermined functions, such as spline functions or polynomial functions. 
For instance, \citet{ramm1993shape} proposed a shape optimization for shell structures using B\`{e}zier patches to represent the shape. \citet{jiang2021shape} employed B-spline surface representation to optimize static and dynamic behaviors of shells. 
Parametric methods confine the allowable shape changes, although the optimization is robust by ensuring a smooth surface. 

In contrast, the nonparametric methods can represent freeform shapes that are not based on a predetermined function. Typically, changes in shape are treated as offsets of each mesh node. \citet{firl2013regularization} studied a nonparametric shell shape optimization by a filter and regularization scheme for meshes. \citet{shimoda2014non} proposed a nonparametric method for freeform shell shape optimization. To ensure the smoothness of structure, the design is updated to the deformed state when the shape sensitivity is corresponded to a traction force. 

Although nonparametric methods can represent freeform shapes, they cannot handle large design changes in shape and topology of the shells. The primary reason for this challenge is that the mesh is distorted as the shape changes from its initial state. When the mesh distortion becomes intolerably large, remeshing is an intuitive solution. However, it is a rather challenging problem to regenerate the whole mesh while maintaining the smoothness of the shape. Handling topological changes will be even more challenging. Moreover, a self-intersection of the shells cannot be resolved because the shell analysis assumes no connection at the intersection, which cannot be physically realized.

Topology optimization on the shells has also been studied. \cite{park2008topology} proposed a level set-based topology optimization method for a material distribution problem on shells. \cite{huo2022topology} applied a moving morphable method for topology optimization on complex surfaces using conformal mapping. It is often combined with shape optimization, both parametric methods~\citep{ansola2002integrated,hassani2013simultaneous,kang2016isogeometric,jiang2023explicit} and nonparametric methods~\citep{shimoda2021unified,tan2022efficient}. Simultaneous optimization of shape and topology has gained attention in recent years as it can increase design freedom by considering material distribution on the surface. However, both parametric and nonparametric shape optimization in their methods still suffer from the aforementioned problems for significant design changes of the surface itself. 

To overcome these issues, we propose a novel method for optimizing shell structures with large design changes in shape and topology. The key idea of our method is to introduce a level set function for representing the midsurface of the shell by the zero isosurface of the level set function. Arbitrary manifold shapes can be represented by the level set function in a relatively coarse volume mesh. Shell analyses are performed on the evolving surface mesh. To extract the surface mesh, we utilize an open source remeshing software Mmg~\citep{mmg}. The shape sensitivity can be mapped to level set function sensitivity by introducing assumptions from a discussion on the profile of level set function around the zero isosurface. We verify the utility of the proposed method by applying to the mean compliance minimization of dome, plate and cantilever beam 3D shell design problems. 

The rest of this paper is organized as follows. In Section \ref{sec_2_formulation}, formulation of shell representation, shell analysis, and optimization problem are presented. In Section \ref{sec_3_implementation}, implementation is presented, including mesh generation and sensitivity information mapping from surface to volume mesh. In addition, the overall optimization process is discussed. In Section \ref{sec_4_numerical_examples}, numerical examples are provided. The dome example verified that an intuitive optimized solution can be obtained. The plate and cantilever beam problems demonstrated the proposed framework can handle large design changes in shape and topology. Finally, Section \ref{sec_5_conclusion} concludes this study.

\section{Formulation}\label{sec_2_formulation}
\subsection{Representation of shell structure based on level set function}\label{ssec_rep}
\begin{figure}[bp]
    \centering
    \includegraphics[width=70mm]{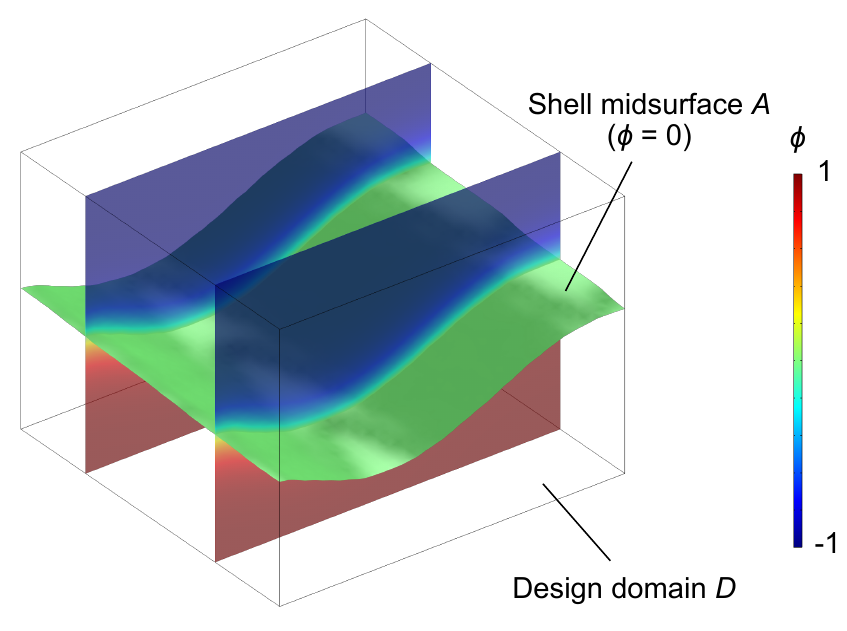}
    \caption{Schematic of design domain $D$, level set function $\phi$, and midsurface of shell $A$\label{fig_schematic}}
\end{figure}

We consider a level set function $\phi$ in a three-dimensional design domain $D$. The midsurface of the shell $A$ is represented by the zero isosurface of the level set function $\phi$, as follows:
\begin{align}
\phi(\bm{x}) = 0 \quad \text{for} \quad \forall \bm{x} \in A, \label{eq_lsf}
\end{align}
where $\bm{x}$ is the position vector in the design domain $D$. 
Figure~\ref{fig_schematic} shows the schematic of the design domain $D$, level set function $\phi$, and midsurface $A$. We define the shell structure as a structure having a uniform thickness $t$ normal to the midsurface $A$, i.e., $\Omega = A \times (-t/2,\ t/2) \subset {\mathbb{R}}^{3}$.

We can represent arbitrary manifold shell shapes by introducing this level set function. Unlike conventional shell shape optimization, the level set-based representations can naturally handle the intersection of two adjacent surfaces by generating holes. For example, when the shape changes such that the spherical surface is pinched, it maintains the manifold shape by becoming a torus. Therefore, this method can express topology changes and large shape changes. 

\subsection{Filtering and projection scheme}\label{ssec_filter_proj}
The level set function $\phi$ must be sufficiently smooth to ensure the continuity of the midsurface $A$. In addition, $\phi$ should have constant gradient $\| \frac{\partial \phi}{\partial \mathbf{x}} \|$ around $\phi = 0$ so that the shape sensitivity on the shell can be associated with the sensitivity of the level set function, as discussed later in Section~\ref{ssec_update}. For these reasons, we introduce filtering and projection techniques.

To ensure the spacial smoothness of the level set function, we introduce a Helmholtz-type filter \citep{kawamoto2011heaviside,lazarov2011filters} as follows:
\begin{align}
    -R^2 \nabla^2 \tilde{\psi} + \tilde{\psi} = {\psi}\quad \mathrm{in}\ D,\label{eq_helmholtz}\\
    \nabla \tilde{\psi} \cdot \mathbf{n} = 0\quad \mathrm{on}\ \partial D,
    \label{eq_phi_neumann}
\end{align}
where $R$ is the filtering radius, $\mathbf{n}$ is the normal vector on design domain boundaries, and $\psi$ is a scalar function. $\psi$ is the design variable in the context of the optimization problem. The filter radius $R$ controls the minimum length scale in the level set function. Accordingly, a larger $R$ leads to a smaller maximum curvature of shells. In the proposed method, a large enough $R$ is needed to ensure that the curvature of shells can be represented by the actual mesh size.

For the projection function, a regularized Heaviside function \citep{kawamoto2011heaviside} is used:
\begin{equation}
    \hat{\phi} = 
    \begin{cases}
        -1 & \quad \mathrm{for}\ \tilde{\psi} < -h\\
        2H(\tilde{\psi}) - 1 & \quad \mathrm{for}\ -h \leq \tilde{\psi} \leq h\quad ,\\
        1 & \quad \mathrm{for}\ \tilde{\psi} > h
    \end{cases}
    \label{eq_heaviside}
\end{equation}
where $h$ is a positive parameter for the bandwidth. $H(\tilde{\psi})$ is defined as:
\begin{equation}
    H \left(  \tilde{\psi} \right) = \frac{1}{2} + \frac{15}{16} \left( \frac{ \tilde{\psi}}{ h}\right) - \frac{5}{8} \left( \frac{ \tilde{\psi}}{ h}\right)^3 + \frac{3}{16} \left( \frac{ \tilde{\psi}}{ h}\right)^5.
    \label{eq_poly}
\end{equation}
By applying Eq.~(\ref{eq_heaviside}), binarization of $\hat{\phi}$ will be enhanced. After that, we apply the same type of filter as in Eqs.~(\ref{eq_helmholtz})--(\ref{eq_phi_neumann}) to obtain the level set function $\phi$ while having a nearly constant absolute value of the gradient 
$\| \nabla \phi \|$
around $\phi = 0$:
\begin{align}
    -R^2 \nabla^2 \phi + \phi = {\hat{\phi}}\quad \mathrm{in}\ D,\label{eq_helmholtz2}\\
    \phi = \phi_\mathrm{init}\quad \mathrm{on}\ \Gamma_{\mathrm{D}},\label{eq_phi_dbc}\\
    \nabla \phi \cdot \mathbf{n} = 0\quad \mathrm{on}\ \partial D \backslash \Gamma_{\mathrm{D}}.
    \label{eq_phi_neumann2}
\end{align}
Eq.~(\ref{eq_phi_dbc}) is applied not to change the structure on boundaries $\Gamma_\mathrm{D}$, where $\phi_\mathrm{init}$ is the level set function of the initial structure. Unlike general topology optimization using volume meshes, the proposed method cannot proceed the optimization if the structure disappears on the load boundaries or fixed boundaries because only the shell domain is included for the finite element analysis. We address this issue by applying Eq.~(\ref{eq_phi_dbc}) to maintain an initial structure on specific boundaries. 

\subsection{Design update of shell structure}\label{ssec_update}
We introduced the level set function for representing the shell structure as mentioned in Section~\ref{ssec_rep}. For the optimization, the level set function defined on a volume mesh has to be updated instead of updating the shell surface itself. However, the shell analysis provides the shape sensitivity of the shell. Therefore, it is necessary to figure out the relationship between the shape sensitivity of the shell and the sensitivity of the level set function. 

\begin{figure}[tbp]
    \centering
    \includegraphics[width=70mm]{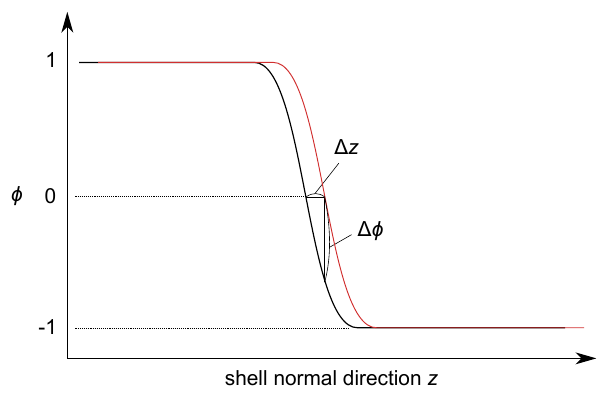}
    \caption{1D cut plot of the level set function in the normal direction of the gradient\label{fig_shell_ls}}
\end{figure}

As a simple example, let us consider a 1D cut plot shown in Fig.~\ref{fig_shell_ls}. The level set function transitions from $-1$ to $1$ around the domain where the shell exists. The horizontal axis $z$ in Fig.~\ref{fig_shell_ls} represents the local coordinate normal to the shell. The direction in which the function decreases is defined as the positive direction. To associate the shell shape sensitivity in normal direction $\frac{\partial F}{\partial z}$ with the sensitivity of the level set function $\frac{\partial F}{\partial \phi}$, the relationship between a finite difference $\Delta z$ and $\Delta \phi$ should be determined. Here $F$ denotes the objective function. Assuming that $\Delta z$ is smaller than the bandwidth of the level set function, $\frac{\Delta \phi}{\Delta z}$ can be approximated by the absolute of the spacial gradient $\| \nabla \phi \|$ because the spacial gradient is nearly constant around $\phi = 0$ region. 

Based on the above discussion, we introduce an approximated sensitivity $\widehat{\frac{\partial F}{\partial \phi}}$ as follows:
\begin{equation}
    \widehat{\frac{\partial F}{\partial \phi}} = - c \frac{\partial F}{\partial z},
    \label{eq_shell_ls}
\end{equation}
where $c$ is a constant that should be set to representative value of $1 / \| \nabla \phi \|$.

The approximation error of the sensitivity is composed of several factors, such as surface mesh extraction and sensitivity projection from the surface mesh to the fixed mesh, and affects the accuracy of the design update. One important factor for the accuracy of the design update is the gradient of the scalar field near the zero-value boundary.

In the proposed method, the spatial gradient $\nabla \phi$ at $\phi=0$ is implicitly controlled by a combination of filtering and projection (Eq.~(\ref{eq_helmholtz})--(\ref{eq_phi_neumann2})); however, the uniformity of the spatial gradient is not guaranteed mainly because it depends on the geometry, e.g., curvature and neighboring distance of zero isosurface. If the actual spatial gradient $\nabla \phi$ were used in Eq.~(\ref{eq_shell_ls}), the level set sensitivity derived from the shape sensitivity would be extremely large when $\| \nabla \phi \|$ is very small, and the design collapses by update. Assuming $c$ as constant ensures that the optimization remains stable even with large shape changes and topological changes where the linear expansion of the level set function may not hold, although the design update is not proportional to the shape sensitivity.

\subsection{Shell analysis}\label{ssec_shell_analysis}
The variational formulation for linear elastic shells, assuming plane stress condition where $\sigma^{33}=0$, is given by
\begin{align}
\label{eq_constitutive}
 &     \int_{\Omega} \left[ \mathscr{C}^{\alpha\beta\lambda\mu} \right. \epsilon_{\alpha\beta}(\bm{u})   \epsilon_{\lambda\mu}(\bm{v})   \nonumber \\ &+ \frac{5}{6}\left. \mathscr{D}^{\alpha\lambda} \epsilon_{\alpha 3}(\bm{u})   \epsilon_{\lambda 3}(\bm{v}) \right] d\Omega
     = 
     \int_{\Omega} \bm{f} \cdot \bm{v}\ d\Omega,
\end{align}
where the indices $\alpha$, $\beta$, $\lambda$ and $\mu$ range from 1 to 2, and superscripts represent contravariant indices, whereas the subscripts denote covariant indices. 
In Eq.\eqref{eq_constitutive}, $\Omega$ refers to the reference domain of the shell structure, $5/6$ is a correction factor for transverse shear deformations, $\epsilon_{\alpha\beta}$ indicates covariant components of the strain tensor, and $\mathscr{C}^{\alpha\beta\lambda\mu}$ and $\mathscr{D}^{\alpha\lambda}$ are contravariant components of the modified constitutive tensors as
\begin{align}
& \mathscr{C}^{\alpha\beta\lambda\mu}= \nonumber\\ 
& \frac{E}{2(1+\nu)}\left( g^{\alpha\lambda} g^{\beta\mu}+g^{\alpha\mu} g^{\beta\lambda}+\frac{2\nu}{1-\nu} g^{\alpha\beta} g^{\lambda\mu} \right),
    \label{modified_constitutive1}
\end{align}
and
\begin{equation}
\mathscr{D}^{\alpha\lambda}= \frac{2E}{2(1+\nu)} g^{\alpha\lambda},
    \label{modified_constitutive2}
\end{equation}
wherein $E$ is Young's modulus, $\nu$ is Poisson's ratio and  $g^{\alpha\beta}$ is contravariant components of metric tensor.
$\bm{f}$ denotes the external 3D loading applied to the shell structures, $\bm{u}$ is the trial function representing the unknown displacements that satisfy the Reissner-Mindlin kinematical hypothesis and boundary conditions,  and $\bm{v}$ is any corresponding test function of $\bm{u}$. 

In this study, computation of shell analyses and their sensitivities is conducted by using MITC6 triangular 6-node shell finite elements implemented in COMSOL Multiphysics \citep{dominguez2018stress}. Note that shell elements other than MITC6 can be used in the proposed method because the level set-based structural representation only requires that the structure can be discretized into surface mesh.
The MITC stands for ``Mixed Interpolation of Tensorial Components''
which refers to a mathematical technique  utilized for alleviating numerical problems such as shear and membrane locking. These issues often arise in conventional shell element formulations \citep{chapelle2010finite}. In the MITC formulation, the transverse shear strains are replaced with assumed shear strains which are derived from an alternative displacement interpolation and are then linked to the displacements at specific tying points. For detailed information on the exact interpolations and the tying points used in MITC6 shell elements, refer to \cite{kim2009triangular, lee2010quadratic}.

The geometry of the MITC6 shell element is interpolated using the following equation:
\begin{equation}
\bm{x}=\sum^6_{I=1} N^I(\xi^1,\xi^2)\bm{x}^I+\frac{\xi^3}{2}\sum^6_{I=1} t^I N^I(\xi^1,\xi^2) \bm{V}^I_n,
    \label{postion vector}
\end{equation}
where the superscript $I$ indicates the node number in the shell element. The local coordinates $\xi^1$ and $\xi^2$ follow the midsurface, while $\xi^3$ represents the coordinate in the normal direction.  $N^I$ is the 2D quadratic interpolation function associated with node $I$. The vector $\bm{x}^I$ denotes the position of node $I$ in the global Cartesian coordinate system. $t^I$ and $\bm{V}^I_n$ correspond to the shell thickness and the director vector at node $I$, respectively.

The displacements $\bm{u}$ of the element are expressed by the following equation:
\begin{align}
\bm{u} &=\sum^6_{I=1} N^I(\xi^1,\xi^2)\bm{u}^I  \nonumber\\
&+ \frac{\xi^3}{2}\sum^6_{I=1} t^I N^I(\xi^1,\xi^2) (-\bm{V}^I_2\theta^I+\bm{V}^I_1\phi^I),
    \label{displacement vector}
\end{align}
where $\bm{u}^I$  represents the displacement degrees of freedom (DoFs) in the global Cartesian coordinate system, $\bm{V}^I_1$ and $\bm{V}^I_2$ are unit vectors orthogonal to $\bm{V}^I_n$ and to each other, and $\theta^I$ and $\phi^I$ refer to the rotational DoFs around $\bm{V}^I_1$ and $\bm{V}^I_2$, respectively.

The covariant strain components are calculated using
\begin{equation}
\epsilon_{\alpha\beta}(\bm{u})=\frac{1}{2}\left( \frac{\partial \bm{u}}{\partial \xi^\alpha}\cdot\frac{\partial \bm{x}}{\partial \xi^\beta}+\frac{\partial \bm{u}}{\partial \xi^\beta}\cdot\frac{\partial \bm{x}}{\partial \xi^\alpha}\right),
    \label{inplane strain components}
\end{equation}
for in-plane strains, and
\begin{equation}
\epsilon_{\alpha 3}(\bm{u})=\frac{1}{2}\left( \frac{\partial \bm{u}}{\partial \xi^\alpha}\cdot\frac{\partial \bm{x}}{\partial \xi^3}+\frac{\partial \bm{u}}{\partial \xi^3}\cdot\frac{\partial \bm{x}}{\partial \xi^\alpha}\right),
    \label{transversely shear strain components}
\end{equation}
for transverse shear strains.

To calculate the shape sensitivity in the normal direction of the shell surface, we use the offset feature in COMSOL Multiphysics (for details, see documentation provided by~\cite{comsol56offset}).  
The offset distance $z$ is placed from the midsurface along the normal vector $\bm{V}^I_n$, and therefore the offset $z$ is added to the shell thickness in Eq.\eqref{postion vector} as
\begin{align}
\bm{x} &=\sum^6_{I=1} N^I(\xi^1,\xi^2)\bm{x}^I\nonumber\\
& +\frac{\xi^3}{2}\sum^6_{I=1} (t^I+z^I) N^I(\xi^1,\xi^2) \bm{V}^I_n,
    \label{offset}
\end{align}
where $z^I$ is the offset distance at node $I$.
Finally, the shell shape sensitivity in normal direction $\frac{\partial F}{\partial z^I}$ at node $I$ is computed through Eq.\eqref{offset} as
 \begin{equation}
 \frac{\partial F}{\partial z^I}
 = \left.
 \frac{\partial F(\bm{x})}{\partial \bm{x}}
 \cdot\frac{\bm{x}(z^I)}{\partial z^I} \right\rvert_{z^I=0} .
 \label{shell shape sensitivity}
 \end{equation}
 This calculation is carried out using sensitivity function in COMSOL Multiphysics.

\subsection{Optimization problem}
We consider the minimization of the mean compliance under a volume constraint of the domain enclosed by the midsurface, as follows:
\begin{align}
    & \underset{\psi \in [-1, 1]}{\mathrm{minimize}}& & F 
    := 
    \int_{\Omega} W_\mathrm{s}\ d\Omega,
    \label{eq_objective_function}\\
    & \mathrm{subject\ to}& & G := \int_{D} \frac{\hat{\phi}+1}{2}\ dD - G_\mathrm{max} \leq 0,
    \label{eq_constraint_function}
\end{align}
where $W_\mathrm{s}$ is the elastic strain energy density and $G_\mathrm{max}$ is the prescribed maximum allowable volume. 
Note that we do not apply a typical volume constraint that limits material usage. The proposed shell representation inherently avoids an obvious solution that is entirely filled with material, thus such a material usage constraint is not mandatory.

\section{Implementation}\label{sec_3_implementation}
\subsection{Mesh generation using Mmg}\label{ssec_mesh}
To realize the shell analysis using shell elements, the surface mesh of the shell is required. In addition, it is convenient that nodes have one-to-one correspondence between the surface mesh and the volume mesh when mapping the sensitivity. Hence, we generate a volume mesh conformal to the zero isosurface of the level set function by Mmg, which is an open-source remeshing software~\citep{mmg}. The surface mesh of the shell can be extracted from the conformal volume mesh. 

The level set function value is stored in a regular tetrahedral mesh (hereinafter called ``regular mesh''). Mmg can remesh the regular mesh by providing nodal values of the level set function, resulting a tetrahedral mesh conformal to the zero isosurface of the level set function (hereinafter called ``conformal mesh''). For detailed algorithm for discretizing a surface defined by a level set function, see the article by \cite{dapogny2014three}. 

The mesh size should be enough fine for the filter radius $R$ to accurately represent the level set isosurface and ensure accurate shell analysis. We set the maximum allowed element size by \texttt{-hmax} option in Mmg to keep the mesh resolution.

\subsection{Sensitivity embedding and filtering}\label{ssec_sens_embed}
\begin{figure*}[btp]
    \centering
    \includegraphics[width=160mm]{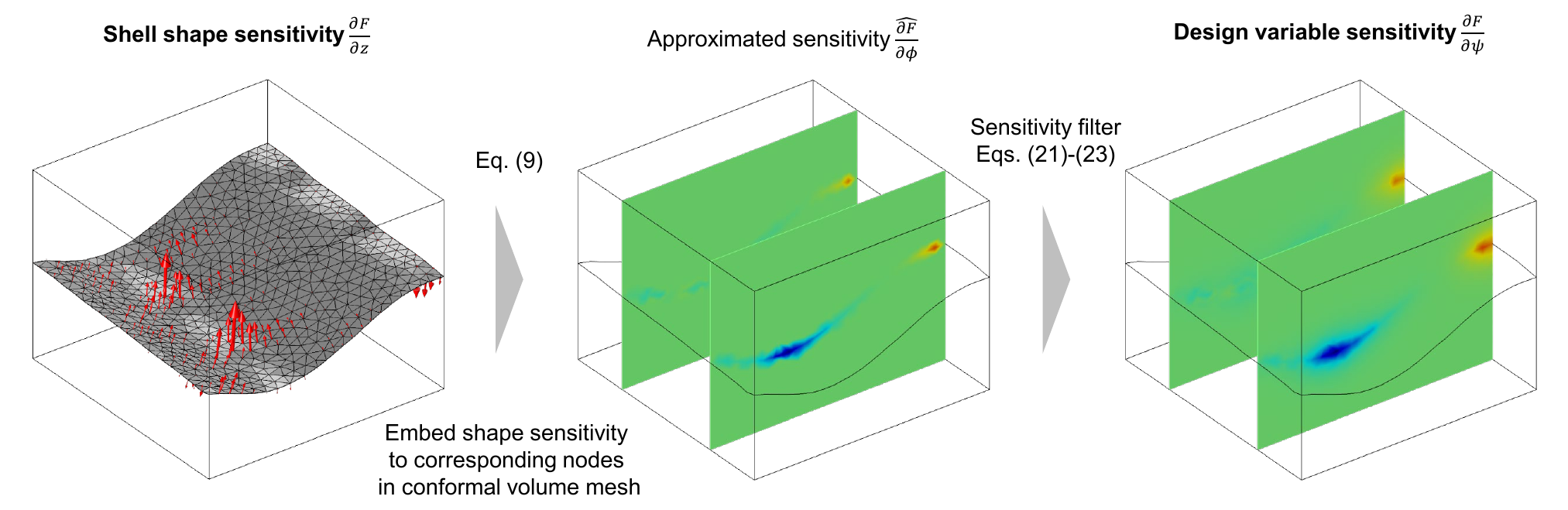}
    \caption{Procedure of design sensitivity derivation\label{fig_sens_embedding}}
\end{figure*}

For the design update of the shell structure, the sensitivity of the level set function given by Eq.~(\ref{eq_shell_ls}) is required. Here, the sensitivity of the original shell offset and level set function are defined by surface and volume, respectively. Thus, the sensitivity field on the surface must be embedded into the volume domain. 
This can be accomplished by embedding the corresponding nodes from surface to volume because the conformal mesh, which is volumetric, contains the nodes of the surface mesh. 
Figure~\ref{fig_sens_embedding} shows the procedure of design sensitivity derivation. 
Note that we allow that the absolute scale of the sensitivity field in volume differs depending on the discretization of the domain, e.g., the mesh size. Yet, the relative value of the sensitivity is consistent to the original shell offset sensitivity.

The spatial smoothness of the sensitivity, which is essential for obtaining a smooth shape, must be ensured. Different from topology optimization dealing with material distribution, the filtering effect is eliminated after extracting shells. Due to this characteristic, the smoothness of the sensitivity is not guaranteed. Therefore, we apply the sensitivity filter independently of the level set function filter, as follows:
\begin{align}
    -R_\mathrm{sens}^2 \nabla^2 \widetilde{\frac{\partial F}{\partial \phi}} + \widetilde{\frac{\partial F}{\partial \phi}} = {\widehat{\frac{\partial F}{\partial \phi}}}\quad \mathrm{in}\ D,\label{eq_sens_helmholtz}\\
    \widetilde{\frac{\partial F}{\partial \phi}} = 0\quad \mathrm{on}\ \Gamma_{\mathrm{D}},\label{eq_dbc_sens}\\
    \nabla \widetilde{\frac{\partial F}{\partial \phi}} \cdot \mathbf{n} = 0\quad \mathrm{on}\ \partial D \backslash \Gamma_{\mathrm{D}},
    \label{eq_sens_neumann}
\end{align}
where $R_\mathrm{sens}$ is the filtering radius. $\widetilde{\frac{\partial F}{\partial \phi}}$ is the smoothed sensitivity and it is assumed to be the sensitivity of design variable $\frac{\partial F}{\partial \psi}$. Since this filter replaces the filter sensitivity calculation in standard topology optimization, which shows the smoothing effect with radius $R$ \citep{kawamoto2011heaviside}, $R_\mathrm{sens}$ should be set comparable to $R$.
Eq.~\eqref{eq_dbc_sens} is introduced to avoid the change of structure on $\Gamma_{\mathrm{D}}$ through the optimization process.

\subsection{Optimization process}
\begin{figure*}[tbp]
    \centering
    \includegraphics[width=140mm]{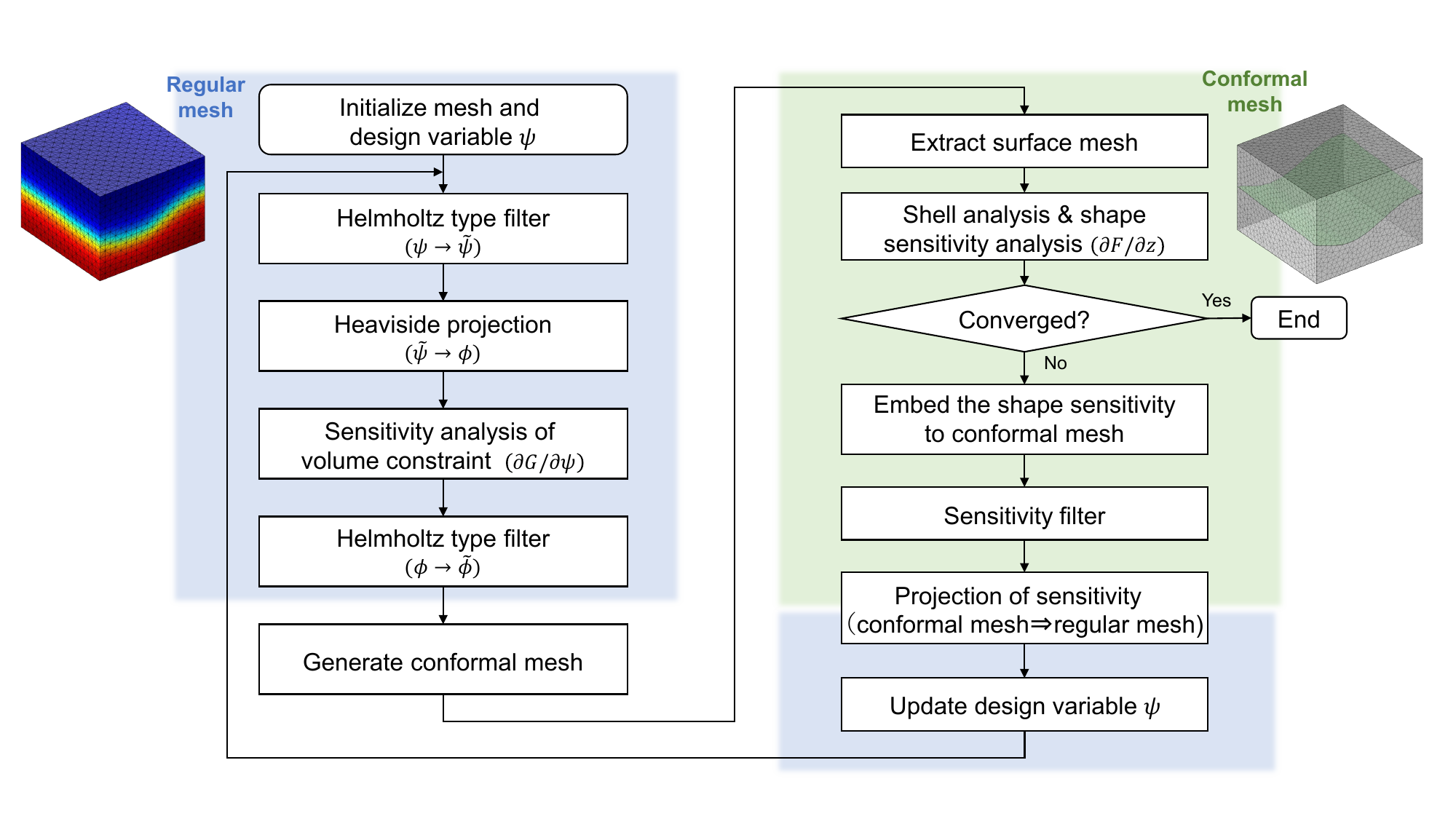}
    \caption{Flowchart of the optimization process\label{fig_flowchart}}
\end{figure*}
The overall optimization process of the proposed method is shown in Fig.~\ref{fig_flowchart}. The filtering and projection procedure is described in Section~\ref{ssec_filter_proj}, and the shell analysis is described in Section~\ref{ssec_shell_analysis}. The latter part, the mesh generation and the sensitivity embedding is explained in Section~\ref{ssec_mesh} and \ref{ssec_sens_embed}. The sensitivity $\frac{\partial F}{\partial \psi}$ is projected from the conformal mesh to the regular mesh using a linear interpolation. This projection allows the optimization process to connect in loop and advance the iteration. The FEM analyses are conducted using COMSOL 5.6, and the other operations including mesh file convert and design variable update are conducted using MATLAB R2020a. 

The steepest descent method is employed as an optimizer because preliminary studies have shown that the solution search becomes unstable if a design update is not proportional to the magnitude of the sensitivity.
Methods such as sequential linear programming, which update at a constant step size based on the sign of the design sensitivity, tend to cause sharp bends in regions where the sensitivity is close to zero. 

We convert the original optimization problem into the unconstrained problem because the steepest descent method cannot handle the constraints directly. The unconstrained optimization problem is formulated using the interior point method~\citep{fiacco1990nonlinear}, as follows:
\begin{align}
    & \underset{\psi \in [-1, 1]}{\mathrm{minimize}}\quad F' 
    := F - \gamma_p \frac{1}{G},
    \label{eq_penalty_function}
\end{align}
where the second term is a barrier function. In the interior point method, the solution is within the feasible region through the optimization process. Since the sign of constraint function $G$ is always negative, the second term in Eq.~(\ref{eq_penalty_function}) becomes larger if the solution is close to the bound of constraint. $\gamma_p$ is a parameter that controls the convexity of the barrier function. We set the initial value of $\gamma_p$ such that the objective function and the constraint function become antagonistic. $\gamma_p$ is decremented each time the solution reaches a stationary point. 

\section{Numerical examples}\label{sec_4_numerical_examples}

\subsection{Dome}
\begin{figure}[tbp]
    \centering
    \includegraphics[width=65mm]{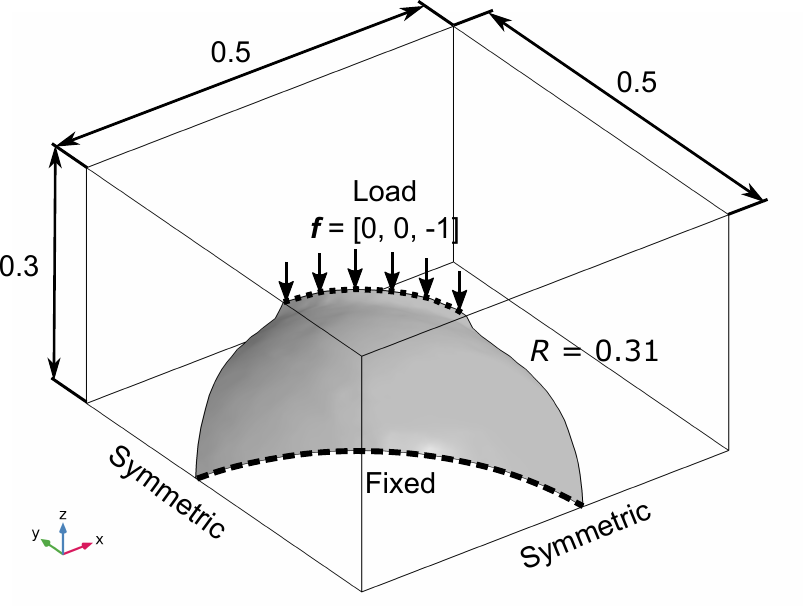}
    \caption{Problem setting of dome example\label{fig_dome_schematic}}
\end{figure}

We demonstrate the application of our proposed method to a dome structure, a common architectural and structural form. The purpose of this example is to show the proposed method can handle typical problems towards evident optimal solutions. Figure \ref{fig_dome_schematic} shows a schematic of the problem setting for this example. Symmetric boundary conditions were applied on $yz$ and $zx$ plane. The size of quarter design domain is $0.5\times 0.5\times 0.3$.

\begin{figure*}[btp]
    \centering
    \includegraphics[width=160mm]{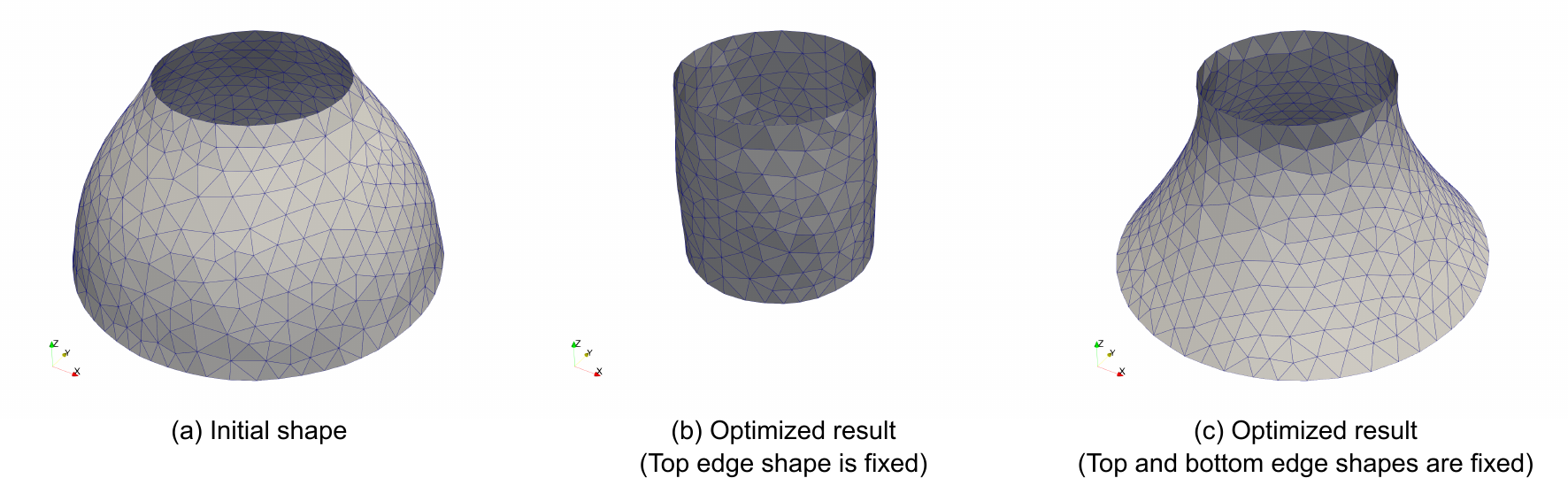}
    \caption{Initial shape and optimized results of dome problem\label{fig_result_dome}}
\end{figure*}
\begin{figure}[tbp]
    \centering
    \includegraphics[width=75mm]{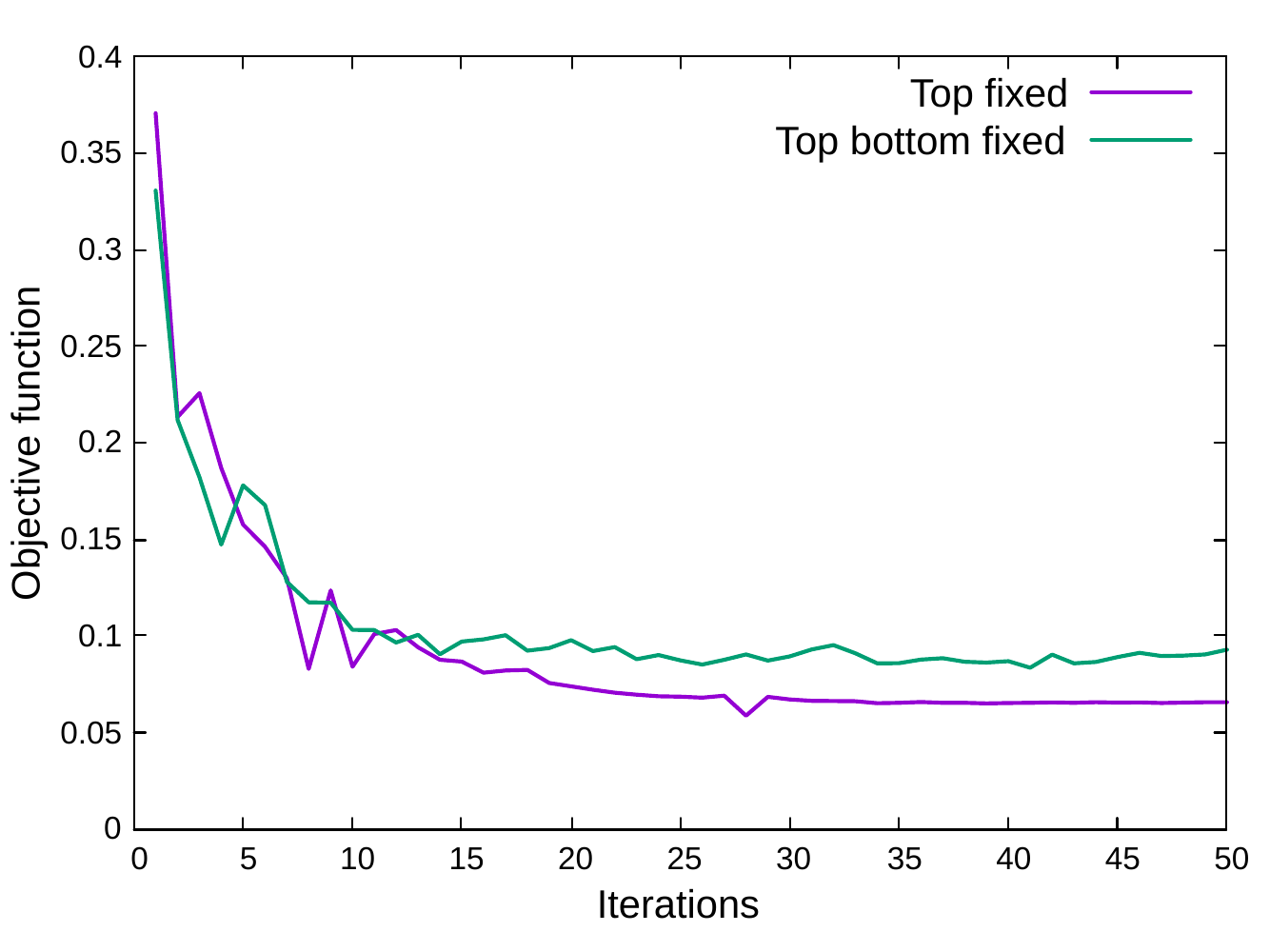}
    \caption{Optimization history of dome problem\label{fig_objhist_dome}}
\end{figure}

We consider two types of geometric constraints in the dome example. In the first case, the optimization proceeds while maintaining the shape of the top edge, allowing changes to the bottom edge. The second case introduces stricter constraints that fix both the top and bottom edges, focusing on internal structural adjustments. The volume constraint is not applied in the dome problem. For both cases, the shell thickness is set to 0.01. The applied load is uniformly vertical and downward, with a magnitude of 1. The Young's modulus is set to $10^5$. The mesh size of regular mesh is 0.05. The maximum and minimum mesh size specified in remeshing are 0.04 and 0.025, respectively. The filtering radius $R$ and $R_\mathrm{sens}$ are set to 0.05. The constant $c$ is set to 15, which is an approximated value of $1 / \| \nabla \phi \|$ in the initial solution. In the optimization process, the update rate of the steepest descent method is set to 100. 

The optimized results are shown in Fig.~\ref{fig_result_dome}. In both cases, the optimization was terminated at 50 optimization iterations because it reached stationary. In the first case (Fig.~\ref{fig_result_dome}(b)), the optimized structure was a cylindrical surface, which is stiffer than the initial structure because it can avoid bending. In the second case (Fig.~\ref{fig_result_dome}(c)), the initial dome structure (Fig.~\ref{fig_result_dome}(a)) was optimized to a concave conical surface. In contrast to the initial structure, which exhibits significant bending around the loaded edge, the optimal structure has a shell surface that transitions smoothly to the fixed edge, which is beneficial to maintain in-plane loading. The objective function was almost monotonically decreasing in the optimization history, as shown in Fig.~\ref{fig_objhist_dome}. The history of the objective function oscillates because the mesh was generated for each iteration. Based on these examples, the validity of proposed method was demonstrated for the typical shell design problems.

\subsection{Plate}
\begin{figure}[tbp]
    \centering
    \includegraphics[width=80mm]{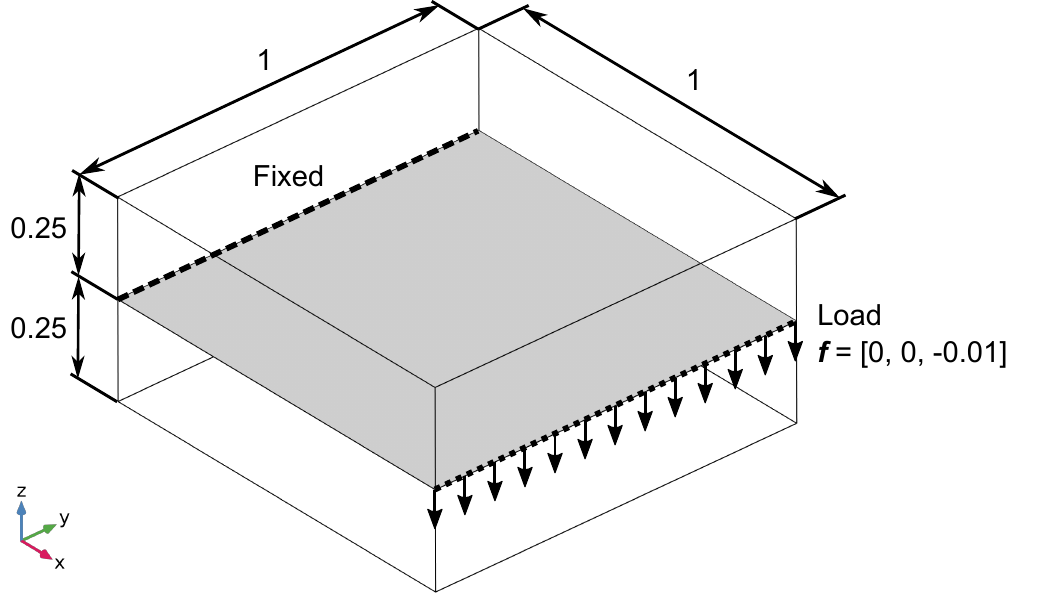}
    \caption{Problem setting of bending plate example\label{fig_plate_schematic}}
\end{figure}

The problem setting for a plate example is shown in Fig.~\ref{fig_plate_schematic}. The initial structure is a flat plate, which shows low stiffness due to low second moment of area. The design domain size is $1\times 1\times 0.5$, where the flat plate is placed at the middle in $z$-axis. The volume constraint and the geometric constraint are not applied to this problem. The applied load is uniformly vertical and downward, with a small magnitude of 0.01 to ensure it complies with the assumption of infinitesimal deformations. The shape of loaded edge can be changed through the optimization process. The Young's modulus is set to $10^5$ and the shell thickness is set to 0.01. The mesh size of regular mesh is 0.05. The maximum and minimum mesh size specified in remeshing are 0.04 and 0.025, respectively. The filtering radius $R$ and $R_\mathrm{sens}$ are set to 0.05. The constant $c$ and the update rate of the steepest descent method are set to 15 and 100, respectively. 

\begin{figure*}[tbp]
    \centering
    \includegraphics[width=160mm]{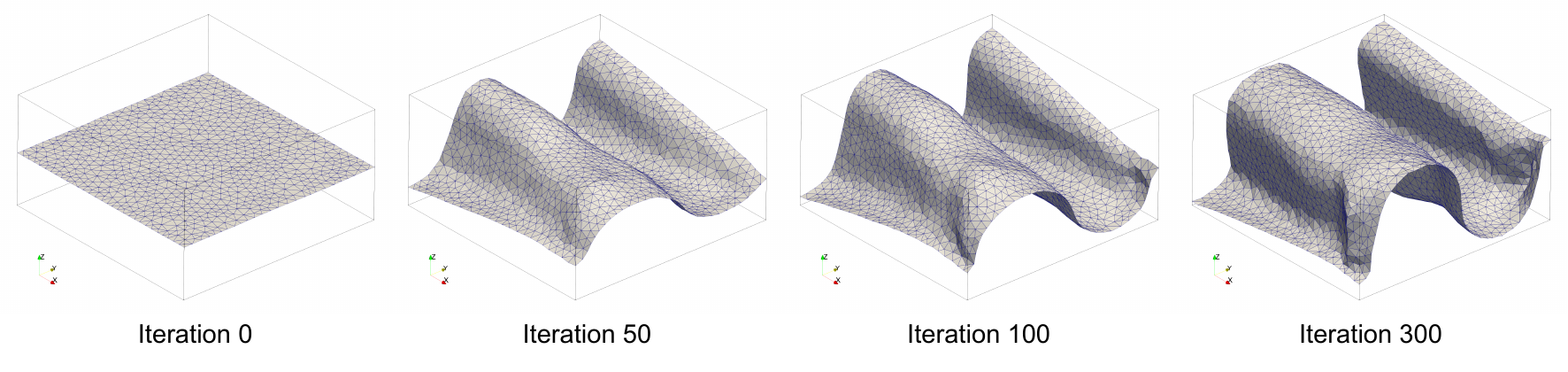}
    \caption{Optimization results of plate problem at different optimization iterations\label{fig_result_plate}}
\end{figure*}

\begin{figure}[tbp]
    \centering
    \includegraphics[width=75mm]{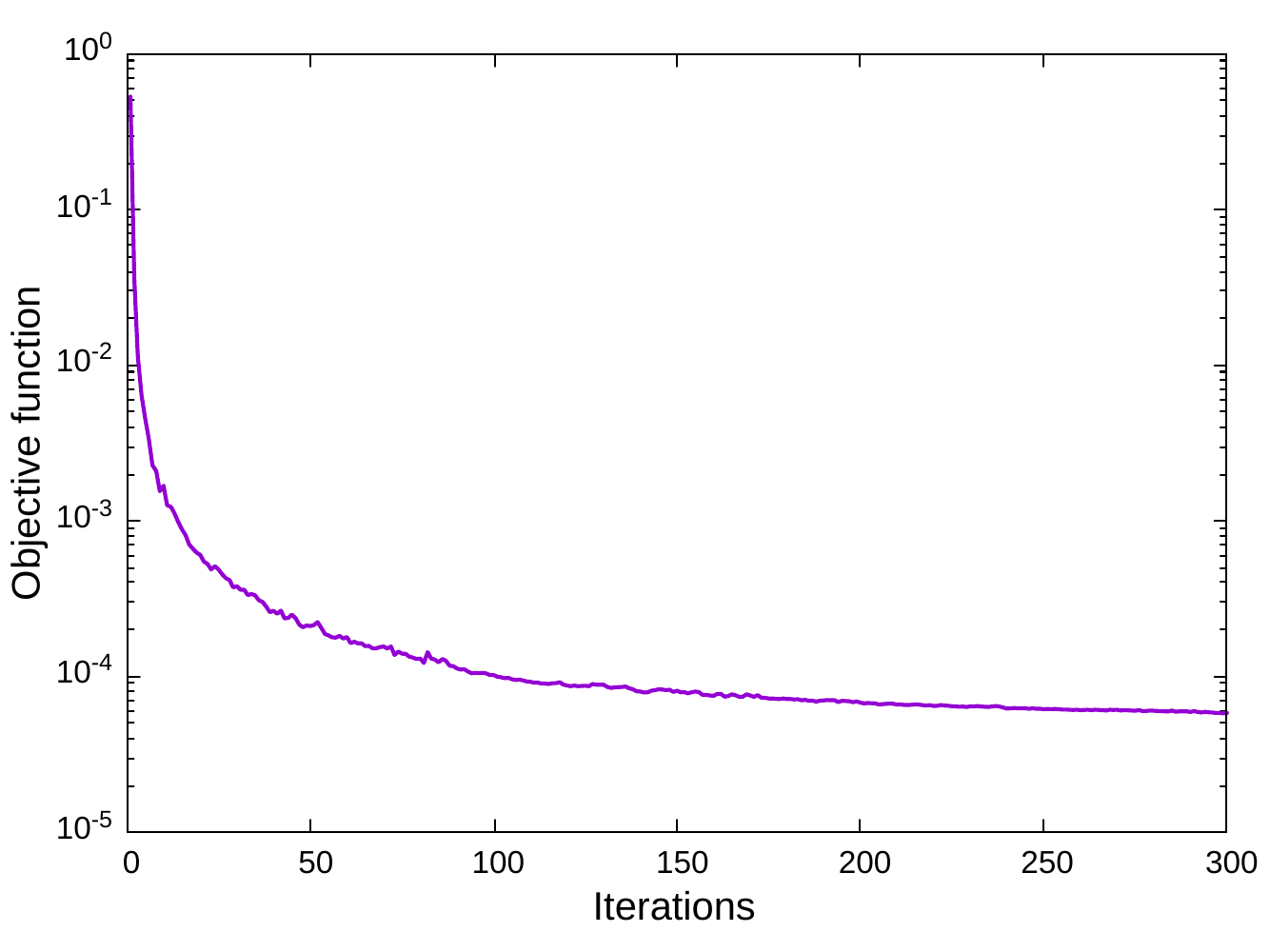}
    \caption{Optimization history of plate problem\label{fig_objhist_plate}}
\end{figure}

The optimization results are shown in Fig.~\ref{fig_result_plate}. The flat plate was gradually changed to a corrugated plate which uses the height of design domain as much as possible. As a result, the mean compliance was significantly decreased as shown in Fig.~\ref{fig_objhist_plate}. 

The corrugated plate configuration demonstrates an increased stiffness against bending loads, consistent with the findings of previous studies on shell shape optimization \citep{shimoda2014non}. The proposed method, which can accommodate larger shape changes, has shown the potential to further improve bending stiffness. Compared to the existing nonparametric methods, the proposed method is advantageous to maintain the mesh regularity through the optimization with large shape changes. The surface mesh at 300 iteration is not distorted in spite of the significant change from initial shape. These results indicate that the proposed method is suitable for the optimization with large shape changes. 

\subsection{Cantilever beam}
\begin{figure}[htbp]
    \centering
    \includegraphics[width=70mm]{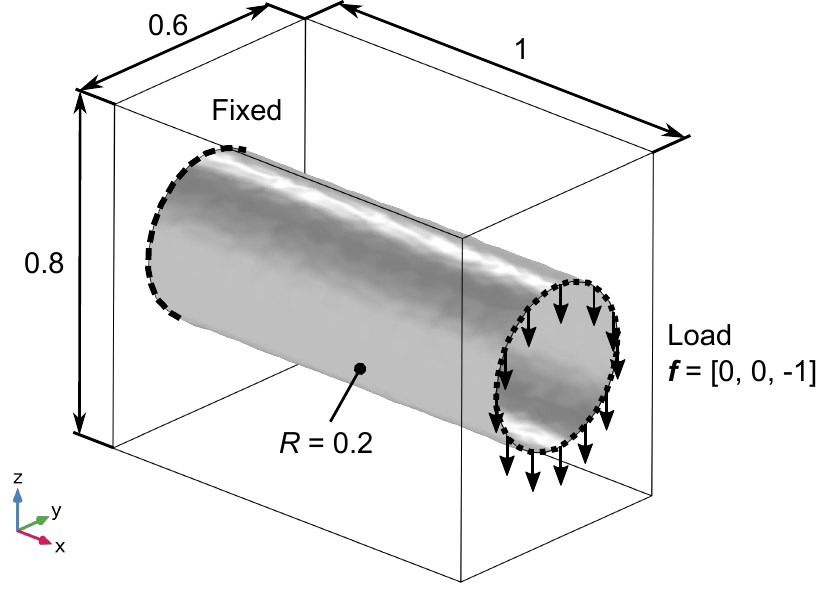}
    \caption{Problem setting of cantilever beam\label{fig_cantilever_schematic}}
\end{figure}

The problem setting for a cantilever beam example is shown in Fig.~\ref{fig_cantilever_schematic}. The initial structure is a cylindrical surface with the radius of 0.2. One end is fixed, and another end is loaded uniformly vertical and downward, with a magnitude of 1. The design domain size is $1\times 0.6 \times 0.8$ where the cylindrical surface is placed at the center in $yz$ plane. The shape change of loaded edge is suppressed by a Dirichlet boundary condition. The Young's modulus is set to $10^5$ and the shell thickness is set to 0.01. The mesh size of regular mesh is 0.05. The maximum and minimum mesh size specified in remeshing are 0.04 and 0.025, respectively. The filtering radius $R$ and $R_\mathrm{sens}$ are set to 0.05. The constant $c$ and the update rate of the steepest descent method are set to 15 and 500, respectively.

In this example, the volume enclosed by the shell is constrained with $G_\mathrm{max}=0.13$ to demonstrate the capability for handling geometric constraints. For the volume constraint, the initial parameter in barrier function $\gamma_p$ was set as follows:
\begin{equation}
    \gamma_{p, \mathrm{init}} = G_{\mathrm{init}}^2 \frac{\|\frac{\partial F_{\mathrm{init}}}{\partial \psi_I}\|_1}{\| \frac{\partial G_{\mathrm{init}}}{\partial \psi_I} \|_1},
\end{equation}
where $F_{\mathrm{init}}$ and $G_{\mathrm{init}}$ are the objective and constraint function value for the initial solution, respectively. The value of $\gamma_p$ was halved when the change rate of $F'$ was less than 0.001 and $G$ was less than -0.01. 

\begin{figure*}[bhtp]
    \centering
    \includegraphics[width=160mm]{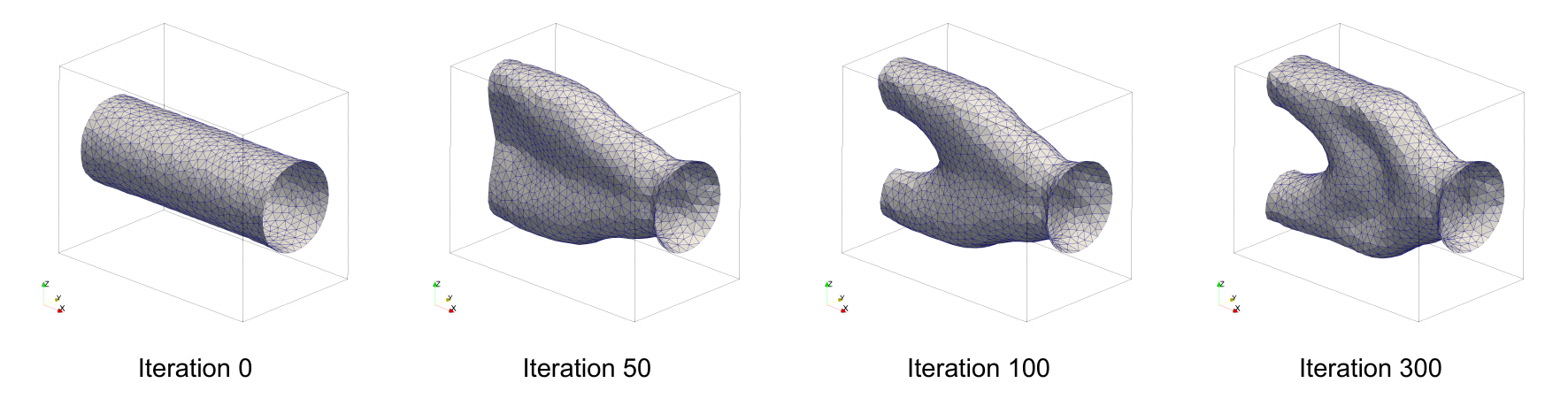}
    \caption{Optimization results of cantilever beam problem at different optimization iterations\label{fig_result_canti}}
\end{figure*}

\begin{figure}[tbp]
    \centering
    \includegraphics[width=75mm]{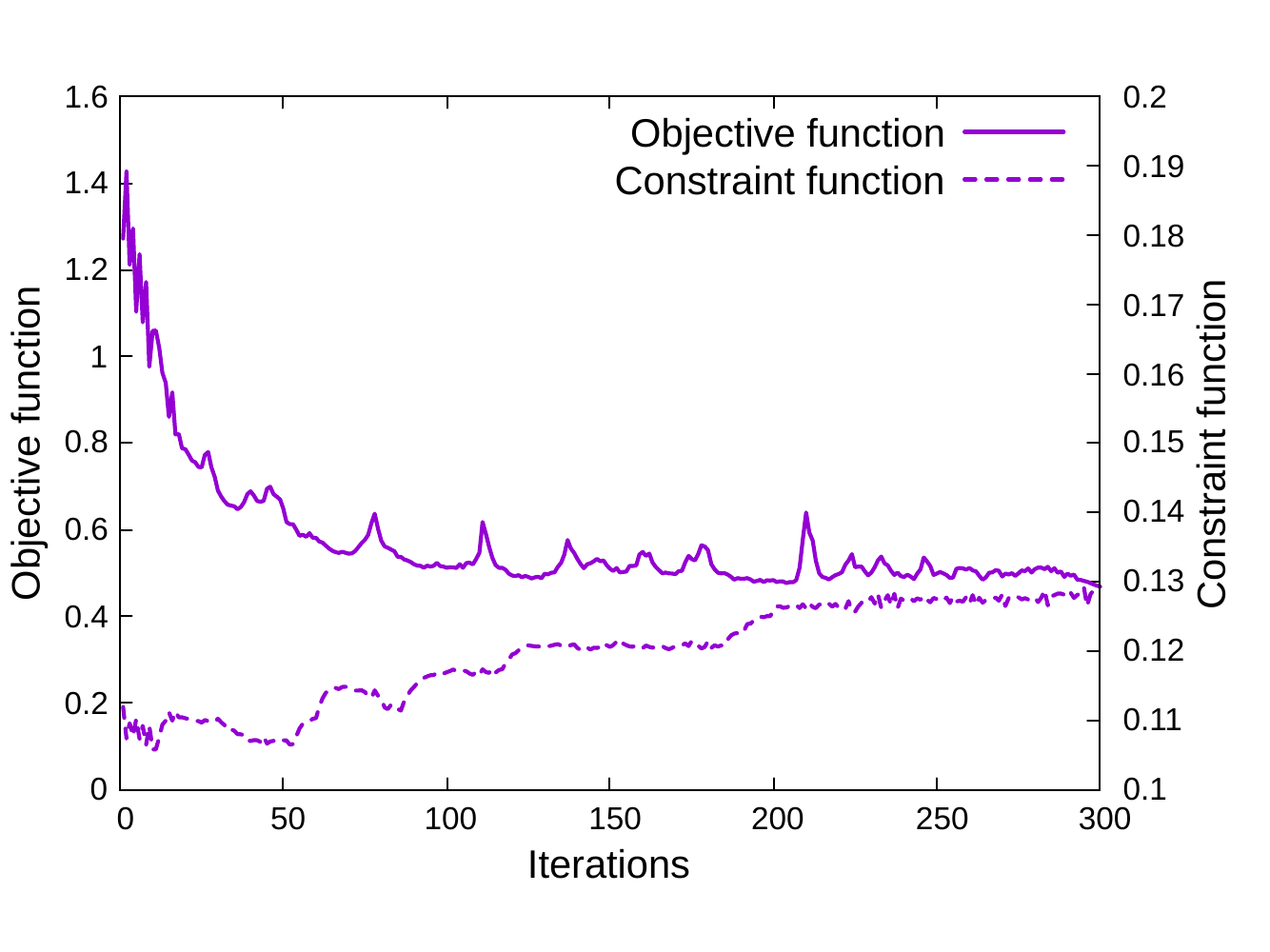}
    \caption{Optimization history of cantilever beam problem\label{fig_objhist_cantilever}}
\end{figure}

The optimization results of the cantilever beam problem are shown in Fig.~\ref{fig_result_canti}. From 0 to 50 iteration, the shape extended in $z$-axis to increase the stiffness. The surfaces around the fixed edge were dented at the center of $z$-axis. From 50 to 100 iteration, the two adjacent surfaces fused together. As a result, a bifurcated structure was obtained from the single cylindrical surface. In other words, the topology of the structure was changed. The optimized structure at iteration 300 has arched ribs on the side, which are often used in sheet metal parts to increase stiffness. As shown in Fig.~\ref{fig_objhist_cantilever}, the optimized structure shows lower mean compliance compared to the initial structure while satisfying the volume constraint of the enclosed domain. The shape change resulting in a thicker root section of the cantilever beam is similar to the vertical section shape in the T-shaped joint result of a previous study \citep{shimoda2021unified}. However, due to the constraint of the enclosed volume in our case, the final shape differed substantially, including the topology.

The cantilever beam example demonstrates the proposed method can deal with large design changes in shape and topology. Such high degree of design freedom has not been achieved by the existing nonparametric shape optimization methods. 

\section{Conclusion}\label{sec_5_conclusion}
This study proposed a level set based optimization method for shell structures with large design changes in shape and topology. We introduced a novel geometric representation, that is, the isosurface of level set function represents the midsurface of the shell. The proposed method involves the conformal mesh generated by remeshing, to bridge the gap between the level set function on the fixed regular mesh and the shell analysis on the surface mesh. We applied the proposed method to dome, plate, and cantilever beam design problems. The optimized results showed that the proposed method can handle large shape changes and topological changes, which is different from the existing nonparametric shape optimization methods. 

The proposed method has an advantage in handling significant structural changes. However, several issues remain. The stability of the method can be compromised by certain filtering and optimization schemes, specifically for large update sizes or updates that are disproportional to the sensitivity, potentially leading to structural collapses. In addition, because our approach relies on local design sensitivity on the shells, it is more prone to local optima than volume mesh-based methods that optimize the entire density field of the design domain. 

For future work, incorporating constraints on the shell material volume, considering buckling effects are promising directions. In addition, addressing material distribution problems on shells and optimizing the layout of stiffeners \citep{jiang2022unified, jiang2023explicit2}, such as ribs, is crucial for industrial applications. Therefore, further advancements in structural representation methods and sophisticated mesh generation techniques are demanded to improve the applicability of the method.

\section*{Funding}
No funding was received for conducting this study.


\section*{Statements and Declarations}
\bmhead{Conflict of interest}
The authors declare that they have no conflict of interest.
\bmhead{Replication of results}
Refer to the Supplementary Movie for the optimization results in iterations. 
The source code is unavailable due to institutional constraints. However, further algorithm details are available upon request to the authors.

\bibliography{sn-bibliography}


\end{document}